# Imaging the proton concentration and mapping the spatial distribution of the electric field of catalytic micropumps

A. Afshar Farniya[1], M.J. Esplandiu[1,2], D. Reguera[3], A. Bachtold[1,4]

[1] ICN2 – Institut Catala de Nanociencia i Nanotecnologia, Campus UAB, 08193 Bellaterra (Barcelona), Spain

[2] CSIC – Consejo Supperior de Investigaciones Científicas, ICN2 Building, Campus UAB, 08193 Bellaterra (Barcelona), Spain

[3] Departament de Física Fonamental, Universitat de Barcelona, C/Martí i Franquès 1, 08028 Barcelona, Spain

[4] ICFO-Institut de Ciencies Fotoniques, Mediterranean Technology Park, 08860 Castelldefels (Barcelona), Spain

Catalytic engines can use hydrogen peroxide as a chemical fuel in order to drive motion at the microscale. The chemo-mechanical actuation is a complex mechanism based on the interrelation between catalytic reactions and electro-hydrodynamics phenomena. We studied catalytic micropumps using fluorescence confocal microscopy to image the concentration of protons in the liquid. In addition, we measured the motion of particles with different charges in order to map the spatial distributions of the electric field, the electrostatic potential and the fluid flow. The combination of these two techniques allows us to contrast the gradient of the concentration of protons against the spatial variation in the electric field. We present numerical simulations that reproduce the experimental results. Our work sheds light on the interrelation between the different processes at work in the chemo-mechanical actuation of catalytic pumps. Our experimental approach could be used to study other electrochemical systems with heterogeneous electrodes.

Catalytic microfabricated engines [1,2] can transform chemical energy into mechanical motion [3,4]. These microengines have generated considerable interest because of the possibility to mimic the fascinating functions of biological motors with man-made engines. Particular attention has been given to self-propelled motors [5-12] and to pumps immobilised on solid surfaces that drive the nearby liquid [13-18]. There has been a growing effort to employ these engines for useful tasks, such as the manipulation of colloidal cargoes, cells, nucleic acids, and bacteria [19-24].

Despite the large number of tasks that have been demonstrated, the mechanism of the chemo-mechanical actuation has been less studied. The actuation mechanism is based on electrochemical processes at the liquid-surface interface of spatially heterogeneous electrodes. It has a lot in common with the physics of basic electrochemical systems [25], corrosion processes [26], energy related devices (such as batteries and fuel cells)[27,28], ion-exchange membranes [29,30], and biological systems (such as biomembranes and ion pumps)[31]. New experimental methods are needed to enable quantitative studies at the microscale of these electrochemical processes.

The actuation mechanism of bimetallic motors/micropumps is based on the oxidation and reduction of hydrogen peroxide ($H_2O_2$) at two different metallic regions acting as anode and cathode, respectively [Fig. 1(a)]. The reactions are [13]

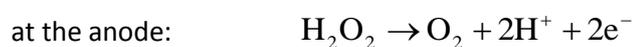

at the anode: $\quad H_2O_2 \rightarrow O_2 + 2H^+ + 2e^-$

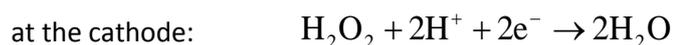

at the cathode: $\quad H_2O_2 + 2H^+ + 2e^- \rightarrow 2H_2O$

Overall, there is a net flux of $H^+$ from the anode to the cathode. The electric field generated in this process is believed to drive motors through electrophoresis and, in the case of micropumps, to induce the flow of the liquid through electro-osmosis [32-37].

The production and the consumption of $H^+$ are related to the electric field in the liquid (and to the electrostatic potential) in an intricate way. On the one hand, the electrochemical reactions, which produce and consume $H^+$, depend, in principle, on the electrostatic potential difference between the liquid and the metal surface as well as on the local concentration of $H_2O_2$ and $H^+$; on the other hand, the rate of the electrochemical reactions controls the current of $H^+$ flowing from the anode to the cathode and, therefore, the variation of the electrostatic potential in the liquid (through Ohm's law).

The interrelation between the catalytic reactions and the electro-hydrodynamics phenomena was analyzed by solving the governing Nernst-Planck, Poisson and Navier-Stokes equations using different approximations [15,32-35]. The spatial variations in the concentration of $H^+$ and in the electric field were found to depend critically on a number of parameters that are difficult to quantify, such as the rate of electrochemical reactions, the zeta potential of metal surfaces, the concentration of ion impurities in the liquid, and their diffusion coefficient. Given the number of ill-defined parameters, and the complexity of the chemo-mechanical actuation, it is important to measure independently the concentration of $H^+$ and the electric field in order to establish the role played by the different processes.

In this Letter, we report on a new method to study the chemo-mechanical actuation of catalytic pumps; it combines two techniques based on optical microscopy. We employed fluorescence confocal microscopy to image and quantify the concentration of $H^+$, a technique used before in biology to measure the local pH [38-41]. The second technique consists in monitoring the velocity in the liquid of particles with different charges in order to map the spatial variations in the electric field, the electrostatic potential, and the fluid flow. Previously, only the magnitudes of the electric field and the fluid flow were estimated [13,15]. The combination of these two techniques allows us to contrast the gradient of the concentration of $H^+$ against the spatial variation in the electric field. It also establishes the zeta potential of metal surfaces. By comparing our experimental findings to numerical simulations, we estimate the concentration of ion impurities and the constant rates of the electrochemical reactions at the anode and cathode. This study provides a quantitative understanding of the chemo-mechanical actuation of catalytic pumps.

Micropumps were fabricated by patterning 30-50 µm diameter platinum disks on gold surfaces using electron-beam lithography and electron-beam evaporation. Platinum and gold were chosen because their electrochemical reactivity in $H_2O_2$ remains rather stable in time [7]. The devices were cleaned with piranha and oxygen plasma to remove residual resist and other organic contaminations. Without this treatment, the motion of the liquid associated to the micropump could not be observed. Electrochemical characterization based on Tafel measurements help to understand this finding (supplementary section III). In absence of the cleaning treatment, the mixed potentials for Au and Pt are very similar. After the treatment, however, the mixed potentials are different. The mixed potential of Pt is at a higher voltage, whereas the mixed potential of Au is at a lower voltage. This is consistent with the measurements discussed below where the platinum disk acts as the cathode, and the gold film as the anode (Fig. 1a). This finding is opposite to what has been observed in gold/platinum motors [5]. This difference may originate from the fact that the cleaning treatment alters the electrochemical properties of the metals by adding some oxygen functionalities to the surface, as demonstrated by X-ray photoelectron spectroscopy measurements (supplementary section II).

We first characterised the electric field and the flow of the liquid [13]. We added positively charged, negatively charged, and quasi-neutral particles to the solution, and then tracked their velocities. As positively charged particles ($p^+$), we used polystyrene spheres functionalized with amidine groups (with a zeta potential $\xi_{p+} = 46$ mV); as negatively charged particles ($p^-$), pristine silica spheres ($\xi_{p-} = -83$ mV); and as quasi-neutral particles ($p^0$), pristine polystyrene spheres ($\xi_{p0} = -12$ mV) (supplementary section I). All the measurements presented in the paper were carried out in an aqueous solution containing 1% $H_2O_2$. Figures 1 b-d show a series of optical images of the motion of a $p^+$ particle. We recorded many movies of the motion of $p^+$, $p^-$, and $p^0$ particles (supplementary section V); Fig. 1f summarises the various types of motion we observed. Particles $p^+$ moved towards the cathode disk, whereas particles $p^-$ did not: they remained more than 20 μm away from its edge. This indicates that the electric field points towards the disk (Fig. 1a). Particles $p^0$ also moved toward the disk; however, once they arrived there, they tended to drift upwards in the direction normal to the disk (Fig. 1f). Since $p^0$ particles interact weakly with the electric field, due to their low charge, their motion reproduces to a good approximation the liquid's flow. The motion of the fluid can be understood as follows [13]: due to electro-osmosis [37], the fluid is driven by the electric field towards the disk in the plane parallel to the surface; it then moves upwards in the direction normal to the disk because of fluid continuity (Fig. 1a).

The spatial variations in the electric field, the electrostatic potential, and the fluid velocity can be estimated from the velocities of $p^+$ and $p^0$ particles measured as a function of the radial coordinate ($r$) along the disk's radius (Fig. 2). The particle velocity has two contributions: one coming from the electrophoretic force ($v_{eof}$) and the other arisen from the fluid flow ($v_f$),

$$v_r = v_{eof} + v_f = \varepsilon \xi E_r / \eta + v_f \qquad (1)$$

where $\varepsilon$ is the fluid permittivity, $\eta$ the fluid viscosity and $\xi$ the zeta potential of the particle. Using Eq. (1) for the velocity of $p^+$ and $p^0$ particles ($v_{p+}$ and $v_{p0}$), the radial component of the electric field and of the fluid flow can be extracted yielding:

$$E_r = \frac{\eta(v_{p+} - v_{p0})}{\varepsilon(\xi_{p+} - \xi_{p0})} \qquad (2)$$

$$v_f = \frac{\xi_{p+} v_{p0} - \xi_{p0} v_{p+}}{\xi_{p+} - \xi_{p0}}. \qquad (3)$$

Figures 3 a-c show the spatial variations in $E_r$, the electrostatic potential ($\phi_r$) (which is obtained by integrating $E_r$), and $v_f$. The electric field and the fluid velocity at the edge of the Pt disk are about 280 V/m and 6 μm s$^{-1}$, respectively. The position at which the electric field approaches zero is consistent with the distance at which the highly negatively charged particles are repelled from the disk. The distance from the disk edge is 28±6 μm [Fig. S5 in supplementary information]. The knowledge of the variation in the electrostatic potential is important (Fig. 3b), because it controls the rate of electrochemical reactions.

We estimate the zeta potential of the substrate ($\xi_w$) from the standard expression of the electrosmotic velocity, $v_f = \varepsilon \xi_w E_r / \eta$. Inserting the data of Fig. 3 a and c for the electric field and the fluid velocity in the previous equation, we obtain a value of the zeta potential that remains nearly constant as a function of the radial distance, with an average value of $\xi_w$= -33 mV. This is close to the values considered for the zeta potential of Au in previous studies [13,15].

We imaged the concentration of $H^+$ using confocal fluorescence microscopy (Fig. 4). For this, we added to the solution a dye molecule (Pyranine) whose fluorescence intensity depends on the concentration of $H^+$. We excited Pyranine at two wavelengths, 405 and 458 nm, and collected the emission from 480 to 580 nm. As the concentration of $H^+$ gets lower, the fluorescence intensity decreases for the excitation at 405 nm, but it increases for the excitation at 458 nm. The ratio between the fluorescence intensities at the two wavelengths allows to measure the local concentration of $H^+$ in the liquid, even if the concentration of the dye is not uniform (due to e.g. the presence of the electric field). The calibration curves and the details of the technique can be found in the supplementary information. Figures 4 a,b show the spatial variation in the concentration of $H^+$. The concentration of $H^+$ is lowest near the Pt disk; it changes by almost one order of magnitude along the radial direction (Fig 4c).

These measurements allow us to contrast the gradient of the concentration of $H^+$ against the spatial variation in the electric field. These results show how $H^+$ ions are produced at the Au anode and move towards the Pt cathode disk where they are consumed. This confirms the direction of the electrochemical reaction and the roles of Au and Pt as anode and cathode, respectively, which was previously inferred from our Tafel plots. The variation in the concentration of $H^+$ near the micropump originates from the chemical reactions involving $H_2O_2$. Indeed, we verified that, in the absence of $H_2O_2$, the $H^+$ ions are homogenously distributed in the liquid (supplementary section VI). We also underscore that the concentration of $H^+$ remains stable over the duration of our experiments (several minutes), indicating that the system reached a steady state.

We now compare our experimental findings to finite elements numerical simulations (supplementary section VII). Previously, an analytical model based on a series of assumptions could account for various facts observed in micropumps [15]. In our work, we directly solved in a coupled way the Poisson equation, the Navier-Stokes equation for the fluid motion, and the Nernst-Planck equation for the transport of species under steady state conditions. We used the zeta potential of the metal surfaces estimated above from our measurements. The rate constants of the electrochemical reactions at the two metal surfaces, and the concentration of ion impurities unavoidably present in the liquid were used as three fitting parameters. Such extra ion impurities could come, for instance, from contamination during the fabrication and the measurement of the micropumps, or the CO2 dissolved in the liquid. Figures 3 a-c and 4c show that the simulations (solid lines) can reproduce both the magnitude and the spatial dependence of the electric field, the fluid flow, the electrostatic potential, and the concentration of $H^+$ in the experiments. The best agreement between simulations and experiments was achieved with the electrochemical rate constants of $k_{Au}$ = 4.1 10$^{-10}$ m s$^{-1}$ and $k_{Pt}$ = 0.01 m$^7$s$^{-1}$ mol$^{-2}$ at the Au and the Pt electrodes, respectively, and a concentration of ion impurities of 1.6 μM. (The units of $k_{Au}$ and $k_{Pt}$ are different, because the current density of protons depends only on the concentration of $H_2O_2$ at the Au surface, whereas it depends on the concentrations of $H_2O_2$ and $H^+$ at the Pt surface.) We emphasize that a concentration of ion impurities in the micromolar range is very low, indicating that contamination from the microfabrication and the measurement setup is weak.

We studied how the 3 fitting parameters affect the electric field, the fluid flow, the electrostatic potential, and the concentration of $H^+$ in the simulations. The rate $k_{Au}$ influences by a large amount all the calculated quantities, whereas $k_{Pt}$ plays comparatively an unimportant role. The concentration of ion impurities $\rho_{imp}$ impacts significantly all the quantities except the concentration of $H^+$. As a result, $k_{Au}$ and $\rho_{imp}$ are the two important fitting parameters; they can be determined in a straightforward way by comparing the data to the simulations. The parameter $k_{Pt}$ is used to fine-tune the fit.

In conclusion, we used two complementary experimental techniques to contrast the gradient of $H^+$ against the electric field of micropumps. Our experimental findings allow us to understand the interrelation between the chemical reactions and the electro-hydrodynamics phenomena of micropumps. Confocal fluorescence microscopy of the concentration of $H^+$ will be a useful tool to investigate other catalytic engines. Indeed, the production and/or the consumption of $H^+$ are central to the chemo-mechanical actuation of many of the catalytic engines fabricated thus far. Our work demonstrates a new approach to probe electrochemical processes at the liquid-surface interface of spatially heterogeneous electrodes. These processes are relevant for a wide range of electrochemical systems.

**Acknowledgements**

We thank I. Pagonabarraga, S. Sanchez, and J. Moser for discussions. We acknowledge M. Roldán, M. de Cabo, and the ''Servei de Microscopia'' of the UAB for support with fluorescence confocal microscopy measurements. We acknowledge support from the European Union (ERC-carbonNEMS project), the Spanish government (FIS2009-11284, FIS2011-22603, MAT2012-31338), and the Catalan government (AGAUR, SGR).

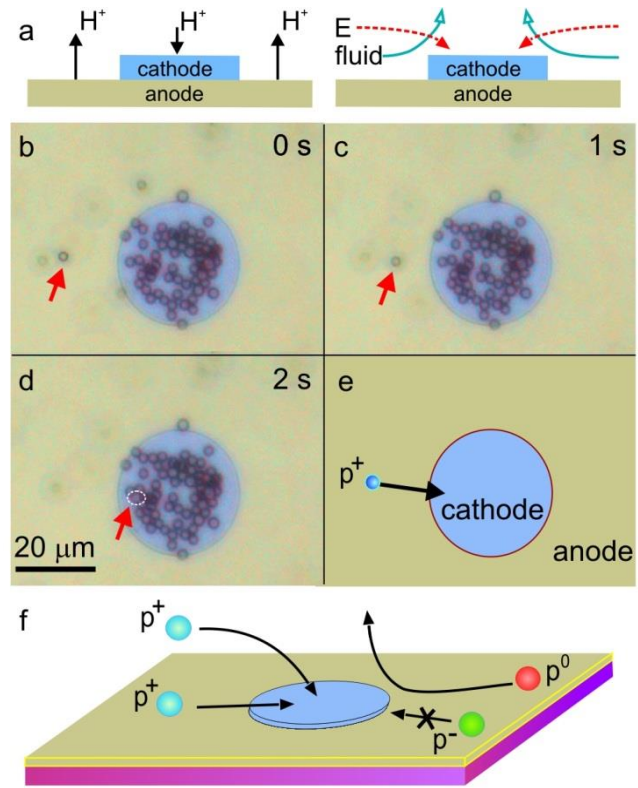

FIG. 1 (a) Schematic representing the production and consumption of $H^+$, the electric field line $E$, and the fluid flow near the micropump. (b-e) Optical images of the motion of a positively charged particle in 1% $H_2O_2$ together with a cartoon depicting schematically its motion. The diameter of the disk is 30 μm. The diameter of the particle is 2 μm. (f) Schematic of the motion of positively charged particles ($p^+$), negatively charged particles ($p^-$), and quasi-neutral particles ($p^0$), based on the measurements of eleven micropumps and 87 particles.

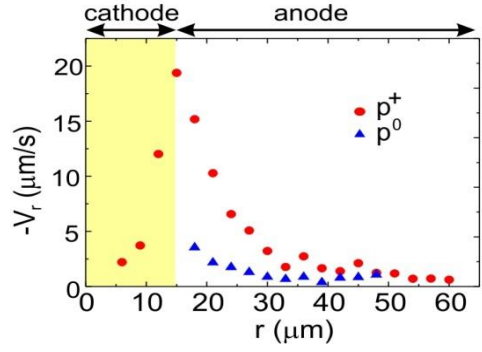

FIG.2 Radial velocity as a function of the radial coordinate $r$. The positively charged particles ($p^+$) and the quasi-neutral particles ($p^0$) are measured using two different micropumps in 1% $H_2O_2$. The diameter of the disks is 30 μm. The origin point is the centre of the cathode disk. The velocity of quasi-neutral particles in the cathode region cannot be recorded, since the particles essentially move perpendicular to the surface or get trapped at the surface near the disk edge (before entering the cathode region). This figure is based on the motion of 25 particles. We studied four micropumps with $p^+$ particles (velocity at each disk edge: 6, 9, 11, and 19 μm·s$^{-1}$); and five micropumps with $p^0$ particles (velocity at each disk edge: 1, 3, 4, 4, and 4 μm·s$^{-1}$). The fluctuations of the velocity from pump to pump might be related to variations in the contamination and the chemical groups at the metal surface. We use the curves of $v_r$ versus $r$ with the highest velocities to estimate the spatial variations in $E_r$ and $v_f$ in Fig. 3, because these curves are less noisy and, thus, better resolve the spatial dependence. Taking the average values of $v_{p+}$ and $v_{p0}$ would reduce $E_r$ and $v_f$ by ~40%.

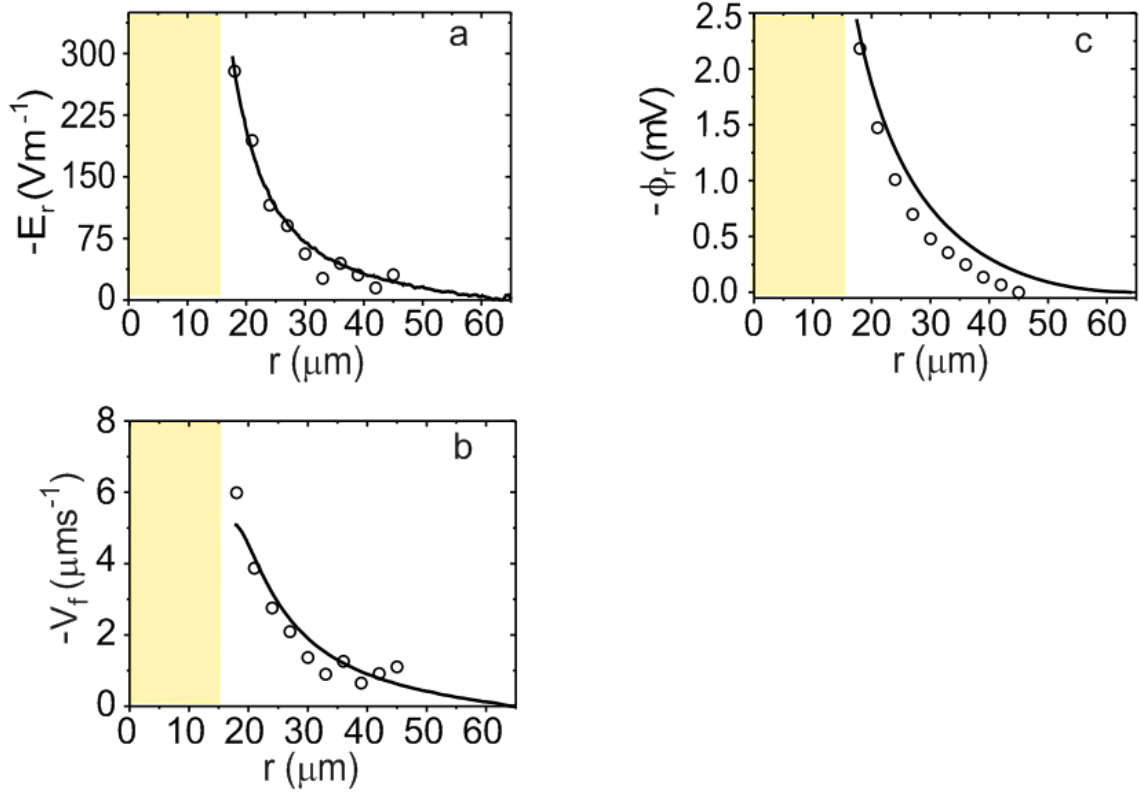

FIG. 3 (a) Radial electric field ($E_r$) plotted against $r$ in 1% $H_2O_2$. $E_r$ is determined from the velocities of $p^+$ and $p^0$ particles in Fig. 2 and using Eq. 2. The diameter of the disk is 30 μm. (b) Electrostatic potential ($\phi_r$) as a function of $r$, obtained by integrating $E_r$. It is natural to define $\phi_r = 0$ for large values of $r$, since the potential of the solution is in equilibrium with that of the metal film and therefore, the electrochemical current is zero. (c) Fluid velocity extracted from Eq. 3. The circles represent the experimental data, whereas the solid lines correspond to the results of the finite elements simulations.

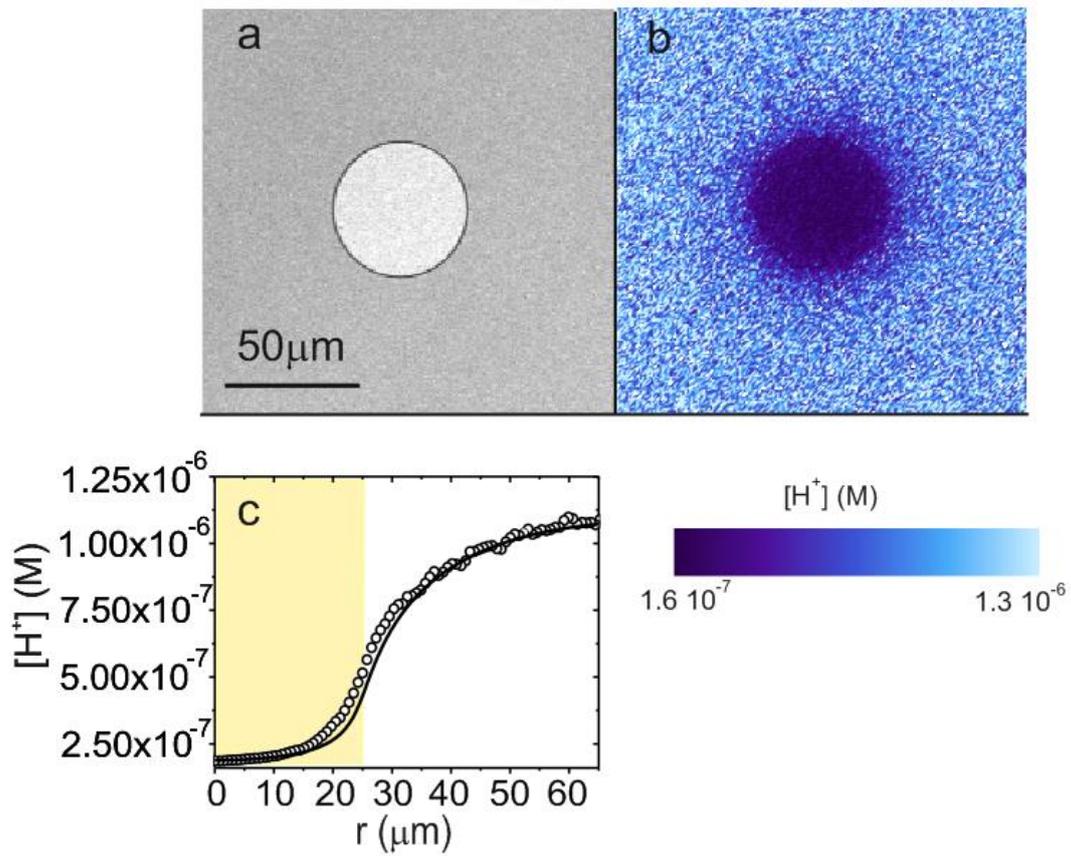

FIG. 4(a) Reflectance image of the micropump obtained with a confocal fluorescence microscope. The diameter of the disk is 50 μm. (b) Image of the concentration of $H^+$. (c) The concentration of $H^+$ as a function of $r$. The plot is obtained from Fig. 4b by averaging the intensity of the pixels along the circle with radius $r$. The circles represent the experimental data, whereas the solid line corresponds to the result of the finite elements simulations.